\documentclass[11pt]{iopart}
\usepackage{iopams}
\usepackage{txfonts}
\usepackage{graphicx}
\begin{document}
\article[Do BCGs have red halos?]{Star-forming Dwarf Galaxies: Ariadne's Thread in the
  Cosmic Labyrinth}{Do blue compact galaxies have red halos?}

\author{E. Zackrisson$^1$, G. Micheva$^1$, N. Bergvall$^2$ and G. \"Ostlin$^1$}

\address{$^1$ Stockholm Observatory, Department of Astronomy, Stockholm University, AlbaNova University Center, 106 91, Stockholm, Sweden}
\address{$^2$ Division of Astronomy \& Space Physics, Uppsala university, Box 515, 751 20 Uppsala, Sweden}
\ead{\mailto{ez@astro.su.se}}

\begin{abstract}
Red halos are faint, extended and extremely red structures that have been reported around various types of galaxies since the mid-1990s. The colours of these halos are too red to be reconciled with any hitherto known type of stellar population, and instead indicative of a very bottom-heavy stellar initial mass function (IMF). Due to the large mass-to-light ratios of such stellar halos, they could contribute substantially to the baryonic masses of galaxies while adding very little to their overall luminosities. The red halos of galaxies therefore constitute potential reservoirs for some of the baryons still missing from inventories in the low-redshift Universe. While most studies of red halos have focused on disk galaxies, a red excess has also been reported in the faint outskirts of blue compact galaxies (BCGs). A bottom-heavy IMF can explain the colours of these structures as well, but due to model degeneracies, stellar populations with standard IMFs and abnormally high metallicities have also been demonstrated to fit the data. Here, we show that due to recent developments in the field of spectral synthesis, the metallicities required in this alternative scenario may be less extreme than previously thought. This suggests that the red excess seen in the outskirts of BCGs may stem from a normal, intermediate-metallicity host galaxy rather than a red halo of the type seen around disk galaxies. The inferred host metallicity does, however, still require the host to be more metal-rich than the gas in the central starburst of BCGs, in contradiction with current simulations of how BCGs form. 
\end{abstract}



\section{Introduction}
Red halos are extended and exceedingly faint structures that have been reported around various types of galaxies since the mid-1990s (see \cite{Zackrisson and Flynn} for a review). The integrated colours of these structures are much too red to be reconciled with any normal type of stellar population, and instead indicative of a halo population with an abnormally high fraction of low-mass stars \cite{Lequeux et al.,Rudy et al.,Zackrisson et al.}. Due to their very high mass-to-light ratios, such stellar populations could contribute substantially to the baryonic masses of galaxies, while contributing no more than a few percent to their overall luminosities \cite{Zackrisson et al.}. The red halos of galaxies therefore constitute potential reservoirs for some of the baryons still missing from inventories in the low-redshift Universe \cite{Fukugita,Nicastro et al.,Prochaska and Tumlinson}. 

While most studies of red halos have focused on disk galaxies (e.g. \cite{Zibetti et al.,Caldwell & Bergvall}), a red excess is also present in the faint outskirts of BCGs \cite{Bergvall & Östlin,Bergvall et al.} -- a type of low-mass, low-metallicity galaxies currently undergoing a dramatic burst of star formation. In this case, however, the reported excess is present in the $K$-band ($\lambda\approx 22000$ \AA) instead of the $i$-band ($\lambda\approx 7700$ \AA) often used in the study of halos of disk galaxies. It may well be that the halos of disks and BCGs have excess light in both $i$ and $K$, but the currently available data do not allow this to be tested. Due to model degeneracies in the near-IR, the BCG halo colours can be interpreted as due to either a bottom-heavy IMF (with the same power-law slope as for the red halos of disks) or due to an unexpectedly high metallicity \cite{Zackrisson et al.}. These degeneracies are not present in the $i$-band -- a high-metallicity population clearly fails to explain the observations of red halos around disks \cite{Zackrisson et al.}. While simplicity would argue for a single explanation (i.e. a bottom-heavy IMF) for the red excess seen around both disk galaxies and BCGs, recent updates in the field of near-IR spectral synthesis modelling calls for a reassessment of the situation. 

\section{The faint outskirts of BCGs}
Deep surface photometry outside the bright star-forming region of BCGs is expected to reveal the starburst progenitor system (host galaxy), and many studies have indeed resulted in the detection of such extended structures around BCGs (e.g. \cite{Noeske et al.,Caon et al.}). However, as demonstrated by Bergvall et al. \cite{Bergvall & Östlin,Bergvall et al.}, the optical/near-IR colours becomes progressively redder as one traces the surface brightness profile outwards, eventually reaching values that are difficult to reconcile with the type of stellar population one naively expects BCG host galaxies to harbour. This prompted Zackrisson et al. \cite{Zackrisson et al.} to suggest that BCGs (and their host galaxies) may be sitting inside red halos similar to those reported around disk galaxies.

\begin{figure}
\centering
\includegraphics[width=10cm]{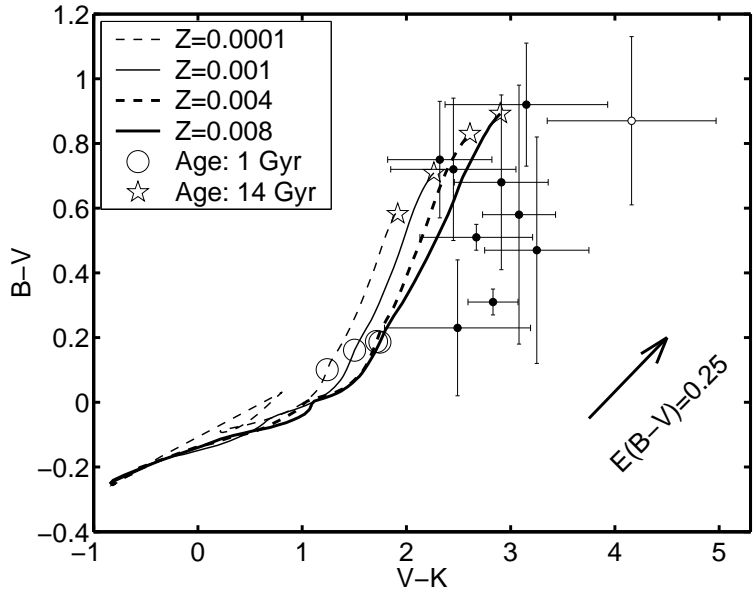}
\caption{Observed $V-K$ and $B-V$ colours of the faint outskirts of BCGs (crosses indicating $1\sigma$ error bars), compared to the predictions of P\'EGASE.2 (lines) for stellar populations with a Salpeter IMF and metallicities spanning the range inferred from emission-line ratios ($Z=0.0001$--0.008. The arrow represents the Milky Way dust reddening vector for $E(B-V)=0.25$. The data points are clearly displaced to the right of the model tracks, indicating a $K$-band excess in the spectra of these objects. Since the dust reddening vector is essentially parallel with the age vector, this displacement cannot be explained by dust extinction. An exponentially declining star formation rate (SFR$(t)\propto \exp(-t/\tau)$) with $\tau=1$ Gyr has here been assumed, but using a different star formation history does not remove the apparent red excess.}
\label{fig1}
\end{figure}

\begin{figure}
\centering
\includegraphics[width=10cm]{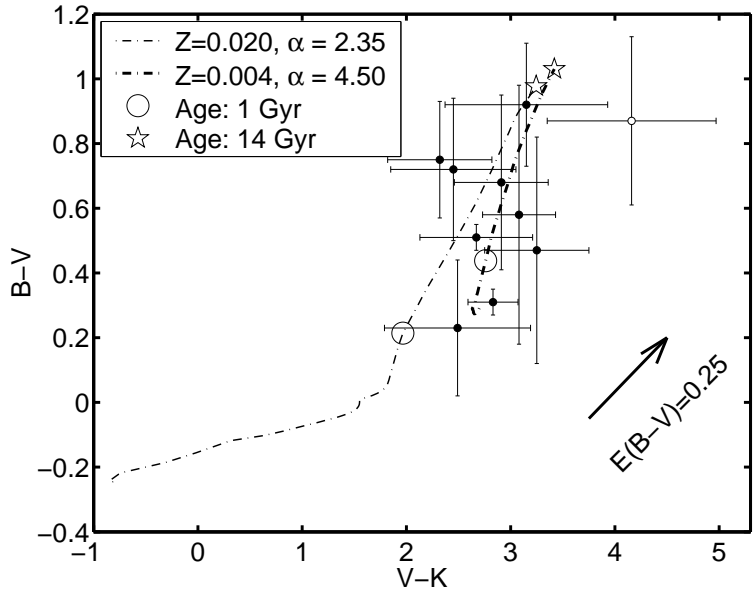}
\caption{Same as Fig.~\ref{fig1}, but showing the P\'EGASE.2 predictions for a Salpeter-IMF population with an abnormally high metallicity ($Z=0.020$, thin dash-dotted line) and a low-metallicity ($Z=0.004$) population with a very bottom-heavy IMF ($dN/dM\propto M^{-\alpha}$ with $\alpha=4.50$, thick dash-dotted line). Contrary to the models shown in Fig.~\ref{fig1}, both of these provide a reasonable fit to the majority of data points after allowing for some dust reddening.}
\label{fig2}
\end{figure}

In Fig.~\ref{fig1} we show the data and model situation for the outskirts of BCGs as it was presented in the Zackrisson et al. paper \cite{Zackrisson et al.}. The gaseous metallicities of BCGs, as measured from emission-line ratios towards the central starburst, are typically around 10\% solar ($Z\approx 0.002$). However, Fig.~\ref{fig1} suggests that no stellar population with a Salpeter-like IMF (including all similar IMFs with breaks around $1 \ M_\odot$) and a metallicitiy in the $Z=0.0001$--0.008 range can explain the observed $B-V$ and $V-K$ colours of the outskirts, regardless of the assumed age or amount of dust reddening. The data points display an offset (of up to $\approx 1$ mag) towards higher $V-K$ with respect to the models, indicating a pronounced $K$-band excess. The model tracks in this diagram are based on the P\'EGASE.2 spectral synthesis code \cite{Fioc & Rocca-Volmerange}, but the majority of models available at the time when the Zackrisson et al. paper \cite{Zackrisson et al.} was written give similar results. As demonstrated in  Fig.~\ref{fig2}, one needs to resort to either a Salpeter-IMF population with a very high metallicity ($Z=0.020$ or even higher) -- in gross conflict with the gaseous metallicities and estimated masses of BCGs -- or a low-metallicity population (here $Z=0.004$) with a very bottom-heavy IMF ($dN/dM\propto M^{-\alpha}$ with $\alpha=4.50$) to explain the data. The latter option implies an abnormally high fraction of low-mass stars and hence a high stellar mass-to-light ratio. Interestingly, Zackrisson et al. \cite{Zackrisson et al.} showed that this combination of metallicity ($Z=0.004$) and IMF slope ($\alpha=4.50$) also explains the red halos detected around stacked disk galaxies in the Sloan Digital Sky Survey \cite{Zibetti et al.}. While we stress that this bottom-heavy IMF solution still remains a viable explanation for the faint outskirts of BCGs, we will demonstrate that due to recent updates in the field of near-IR speactral synthesis, the metallicity required in a Salpeter-IMF scenario is no longer as high as $Z=0.020$.

\section{New model results}
\begin{figure}
\centering
\includegraphics[width=10cm]{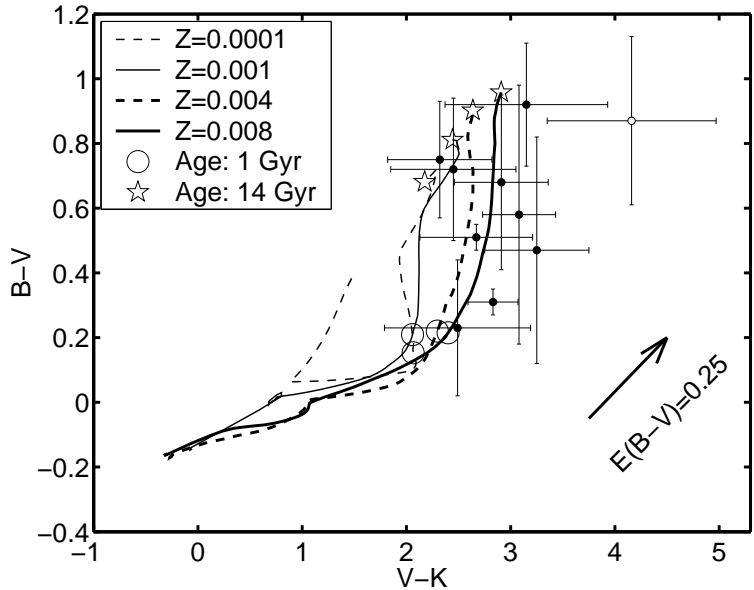}
\caption{Same as Fig.~\ref{fig1}, but with model predictions based on isochrones presented by Marigo et al. \cite{Marigo et al.}. While the predictions are very similar to those of older models (for instance P\'EGASE.2) at high ages in this diagram, the model tracks turn much redder in $V-K$ at intermediate ages (a few Gyrs) due to the updated treatment of TPAGB stars. Because of this, most of the hosts/halos of BCGs can now be readily explained by Salpeter-IMF populations and intermediate metallicities ($Z\leq 0.008$) after allowing for some dust reddening.}
\label{fig3}
\end{figure}

The field of near-IR spectral synthesis has recently seen a small revolution related to the treatment of thermally pulsating asymptotic giant branch (TPAGB) stars \cite{Maraston,Marigo et al.}. While this leaves the interpretation of the $i$-band excess in the red halos of disk galaxies unaffected, it turns out to have a substantial impact on the inferred stellar population properties of the faint outskirts of BCGs, which currently hinges on the interpretation of $K$-band data. In Fig.~\ref{fig3}, we show the predicted evolution of stellar populations with properties (metallicities, IMF and star formation history) identical to those in Fig.~\ref{fig1}, but based on the recent Marigo et al. isochrones \cite{Marigo et al.} instead of the older P\'EGASE.2 model. The most important difference is that the more recent models are significantly redder in $V-K$ at ages around 1 Gyr, allowing the majority of data points to be explained by much lower metallicities ($Z\leq 0.008$) than before, provided that a small amount of dust reddening ($E(B-V)<0.25$) is assumed for the halo region. 

\section{Discussion}
While a bottom-heavy IMF still represents a viable explanation for the optical/near-IR colours of the faint outskirts of BCGs, the most recent spectral synthesis models now suggest that there may be a far more mundane explanation: an intermediate-metallicity stellar population with a perfectly normal IMF. The large error bars on the measured colour make it difficult to assess the exact metallicity requirements for the BCG host galaxy, but close inspection of Fig.~\ref{fig3} indicates a typical metallicity of $Z=0.004$--0.008, unless there is significant dust reddening ($E(B-V)\approx 0.25$) in the outskirts of these systems (which could then lower the required metallicity). While far less extreme than the metallicities of $Z=0.020$--0.040 estimated by Bergvall \& \"Ostlin \cite{Bergvall & Östlin} and Zackrisson et al. \cite{Zackrisson et al.}, this is still a few times higher than the typical metallicities ($Z\approx 0.002$ or $\sim 10$\% solar) measured through emission-line ratios towards the central regions of BCGs. According to the mass-metallicity relations of Panter et al. \cite{Panter et al.}, a metallicity of $Z\approx 0.008$ would be perfectly normal for a stellar mass of $10^{10} \ M_\odot$, but anomalously high for a $10^9 \ M_\odot$ host. The required stellar population masses have not yet been derived for all hosts/halos presented in Figs.~\ref{fig1}--\ref{fig3}, but the analysis of Bergvall \& \"Ostlin \cite{Bergvall & Östlin} indicates that at least some of these objects may have stellar masses around $\sim 10^{10} \ M_\odot$. While this should be looked into on a case-by-case basis, there may no longer be any obvious tension between the host metallicities and inferred masses of BCGs. The one remaining puzzle would then be why the host metallicity seems to be a few times higher than the metallicity of the central starburst ($Z\approx 0.002$). This radial metallicity gradient is in conflict with the simulations of BCG formation presented by Bekki \cite{Bekki} and instead seems to favour a scenario where BCGs form when a low-metallicity gas cloud falls into the centre of an old, intermediate-metallicity host galaxy, thereby igniting a starburst consisting mostly of low-metallicity stars. 

Haro 11 (indicated by an open marker in Figs.~\ref{fig1}--\ref{fig3}), represents a particularly interesting object in the quest to understand the hosts/halos of BCGs, since this is the target with the reddest $V-K$ colours measured so far. While the error bars are admittedly very large, the estimated $V-K\approx 4$ places it well away from the predictions of standard-IMF stellar populations (at any plausible metallicity). If confirmed, this would indicate that there still is something very mysterious about the outskirts of these systems. However, new data for this target (see Micheva et al., these proceedings) suggest that the $V-K$ colour of the host has been overestimated by $\approx 1$ mag. The most likely explanation for the colours observed in the faint outskirts of BCGs therefore seems to be that of a standard-IMF, intermediate-metallicity host galaxy and {\it not} a red halo of the type reported around disk galaxies (e.g. \cite{Zibetti et al.,Caldwell & Bergvall}). Our team is currently in the process of analysing a set of deep $I$-band images of BCGs, which will be able to put this conclusion on a more robust footing.

\ack
EZ acknowledges a research grant from the Royal Swedish Academy of Sciences.


\section*{References}
\begin{thebibliography}{20}
\bibitem{Zackrisson and Flynn} 
Zackrisson, E and Flynn, C 2008, {\it ApJ}, {\bf 687} 242
\bibitem{Lequeux et al.}
Lequeux, J, Fort, B, Dantel-Fort, M, Cuillandre, J-C and Mellier, Y 1996, {\it A\&A} {\bf 312} 1
\bibitem{Rudy et al.}
Rudy, R J, Woodward, C E, Hodge, T, Fairfield, S W and Harker, D 1997, {\it Nature} {\bf 387} 159
\bibitem{Zackrisson et al.}
Zackrisson, E, Bergvall, N, \"Ostlin, G, Micheva, G and Leksell, M 2006, {\it ApJ} {\bf 650} 812
\bibitem{Fukugita}
Fukugita, M 2004, {\it Dark Matter in Galaxies}, ed. S D Ryder, D J Pisano, M A Walker and K C. Freeman (San Francisco: Astronomical Society of the Pacific) p~227
\bibitem{Nicastro et al.}
Nicastro, F et al. 2005, {\it Nature} {\bf 433} 495
\bibitem{Prochaska and Tumlinson}
Prochaska, J X and Tumlinson, J 2008, to appear in {\it Astrophysics in the Next Decade: JWST and Concurrent Facilities}, ed. X. Tielens ({\it Preprint} arXiv:0805.4635)
\bibitem{Zibetti et al.}
Zibetti, S, White, S D M and Brinkmann, J 2004, {\it MNRAS} {\bf 347} 556
\bibitem{Caldwell & Bergvall} 
Caldwell, B and Bergvall, N 2006, in {\it Galaxy Evolution Across the Hubble Time}, ed. F Combes and J Palous (Cambridge: Cambridge University Press), p~82
\bibitem{Bergvall & Östlin}
Bergvall, N and \"Ostlin, G 2002, {\it{A\&A}} {\bf 390} 891
\bibitem{Bergvall et al.}
Bergvall, N, Marquart, T, Persson, C, Zackrisson, E and \"Ostlin, G 2005, {\it Multiwavelength Mapping of Galaxy Formation and Evolution}, ed. A Renzini and R Bender (Berlin: Springer-Verlag, Berlin), p~355
\bibitem{Noeske et al.}
Noeske, K G, Papaderos, P, Cair\'os, L M. and Fricke, K J 2003, {\it A\&A} {\bf 410} 481
\bibitem{Caon et al.}
Caon, N, Cair\'os, L M, Aguerri, J A L and Mu\~noz-Tu\~n\'on, C 2005, {\it ApJS} {\bf 157} 218
\bibitem{Fioc & Rocca-Volmerange}
Fioc, M and Rocca-Volmerange, B 1999, ({\it Preprint} astro-ph/9912179)
\bibitem{Maraston}
Maraston, C 2005, {\it MNRAS}, {\bf 362}, 799
\bibitem{Marigo et al.}
Marigo, P, Girardi, L, Bressan, A¨, Groenewegen, M A T, Silva, L and Granato, G L 2008, {\it A\&A} {\bf 482} 883
\bibitem{Panter et al.}
Panter, B, Jimenez, R, Heavens, A F and Charlot, S 2008, {\it MNRAS} {\bf 391} 1117
\bibitem{Bekki}
Bekki, K 2008, {\it MNRAS} {\bf 388} 10
\end{thebibliography}
\end{document}